# Operationalizing Assurance Cases for Data Scientists: A Showcase of Concepts and Tooling in the Context of Test Data Quality for Machine Learning

Lisa Jöckel[1], Michael Kläs[1], Janek Groß[1], Pascal Gerber[1], Markus Scholz[2], Jonathan Eberle[3], Marc Teschner[3], Daniel Seifert[1], Richard Hawkins[4], John Molloy[4], Jens Ottnad[3]

[1]Fraunhofer Institute for Experimental Software Engineering IESE, Kaiserslautern, Germany
[2]NovelSense, Karlsruhe, Germany
[3]TRUMPF Se + Co. KG, Ditzingen, Germany
[4]University of York, York, UK

[1]{lisa.joeckel, janek.gross, michael.klaes, pascal.gerber, daniel.seifert}@iese.fraunhofer.de
[2]scholz@novelsense.com
[3]{jonathan.eberle, marc.teschner, jens.ottnad}@trumpf.com
[4]{richard.hawkins, john.molloy}@york.ac.uk

**Abstract.** Assurance Cases (ACs) are an established approach in safety engineering to argue quality claims in a structured way. In the context of quality assurance for Machine Learning (ML)-based software components, ACs are also being discussed and appear promising. Tools for operationalizing ACs do exist, yet mainly focus on supporting safety engineers on the system level. However, assuring the quality of an ML component within the system is commonly the responsibility of data scientists, who are usually less familiar with these tools. To address this gap, we propose a framework to support the operationalization of ACs for ML components based on technologies that data scientists use on a daily basis: Python and Jupyter Notebook. Our aim is to make the process of creating ML-related evidence in ACs more effective. Results from the application of the framework, documented through notebooks, can be integrated into existing AC tools. We illustrate the application of the framework on an example excerpt concerned with the quality of the test data.

## 1 Introduction

Assurance Cases (ACs) are a systematic approach to ensure quality in safety engineering. They are defined as "a reasoned and compelling argument, supported by a body of evidence, that a system, service or organization will operate as intended for a defined application in a defined environment" [1]. Most commonly, they are implemented in a tree structure with the quality claim as the root. The claim is iteratively broken down into subclaims until these are modular enough for evidence supporting the claim to be generated.

Quality assurance for software systems with Machine Learning (ML) components is currently a significant area of research and the argumentation of safety and dependability of ML components via ACs is also being discussed [2] [3] [4].

There already exist several tools and frameworks for operationalizing ACs, mainly aimed at supporting safety engineers on the system level [5] [6] [7] [8] [9]. These tools could also be applied for quality assurance of ML components within a software system. However, data scientists are seldom familiar with these tools. Data scientists often use Python and data analysis notebooks like Jupyter Notebook [10], which is a web-based computing environment for usage in a web browser. One of the distinguishing features of Jupyter Notebook is the combination of textual elements, executable code blocks, and computational output, allowing for documents that include textual or visual explanations as well as interaction via executable code. This makes it easier to create reproducible and understandable routines since the code is embedded within the document itself. Notebooks are portable between different users or operating systems and support several programming languages, including those popular with data scientists, such as Python and R.

In this work, we propose the framework pyAC, which is based on Python and Jupyter Notebook. The framework supports data scientists by enabling them to operationalize ACs for ML components in their familiar developing environments. This eliminates the need for them to learn other AC tools in depth. pyAC allows integrating the resulting evidence into external AC tools. This is possible in two ways: (A) First, the claims for the ML component are completely refined in the external tool. Then evidence is generated using pyAC, and finally, the evidence is provided to the external tool. (B) Further refinement of the claims for the ML component is done in pyAC. The evidence is generated for all subclaims and then accumulated evidence is provided to the external tool.

There are three main ***contributions*** in this work. First, we introduce the tooling framework pyAC, which is specifically designed for data scientists to support them in operationalizing ACs for ML components. We present how to integrate this framework into existing AC tools. Second, we describe how the elements and the structure of an AC can be implemented in Jupyter Notebook and Python. Third, we illustrate how to apply the framework on an example AC excerpt where the quality of test data is assured.

The paper is structured as follows: Section 2 provides background and related work on ACs. Section 3 introduces the concept of the tooling framework pyAC. Section 4 illustrates the application of pyAC on test data quality. Section 5 presents future directions and concludes the paper.

## 2     Background and Related Work on Assurance Cases

ACs are an established approach to assure the safety of a software system but are also discussed to assure other qualities, e.g., fairness [11]. They are predominantly implemented in a graphical form as a tree structure, which starts at a *claim* about a system property in its related operating *context*. Based on the iterative application of suitable *strategies*, with their related *assumptions* and *justifications*, a (complex) claim is decomposed into *subclaims* until *evidence* can be provided to justify the validity of the subclaims [12]. ACs can be structured based on, e.g., the Claims-Argument-Evidence (CAE) or the Goal Structuring Notation (GSN) [1] approach, which primarily differ in the designation of their structuring elements [12] [13]. The concept of ACs also appears promising for application in systems containing ML-based components [14] [3] [4].

Various tools supporting the creation of ACs exist [15], such as ASCE [5], ISCaDE [6], Astah GSN [7], from Confiance.ai [8], or safeTbox [9], which are mainly designed to support safety engineers in assuring system-level safety. With these tools, ACs can be created from scratch in a flexible manner for a specific system. Some of them are based on proprietary software platforms, which additionally require a certain amount of expertise. The AMLAS Tool [16] already gives guidance by providing patterns for assuring the ML component that need to be instantiated for a use case. However, to the best of our knowledge, no tools exist that specifically support data scientists in creating evidence for the quality assurance of ML components in a development environment that they use on a daily basis.

For the Open Dependability Exchange (ODE) metamodel, which aims at tool-independent exchange of safety-related artifacts, extensions for enabling the integration of ML assurance-related artifacts are intended [17]. Tools that are compatible with the ODE metamodel, such as safeTbox, could complement pyAC by providing the AC on the comprehensive system level.

## 3  Tooling Framework for Operationalizing Assurance Cases for ML

This section describes the concept of the pyAC framework and how it operationalizes the AC structure and elements using Python and Jupyter Notebook. The framework supports three purposes: (A) It provides guidance for data scientists to implement the AC; (B) it can be applied to a use case to generate evidence for the claimed quality of the ML component; and (C) it can be used for validation of the AC by an (external) assessor, thus providing auditability and reproducibility. An overview of the framework architecture together with the three purposes is shown in Fig. 1.

The framework contains *claims*, *measures*, and *blueprints* as **assurance case elements (AC elements)**, which are instances of their respective Python class (i.e., `Claim`, `Measure`, or `Blueprint`). They inherit from `AssuranceCaseElement`, which provides basic functionalities like storing/loading/deleting a class instance, versioning, or providing an HTML-formatted summary of information on this element. AC elements are created, described, and adjusted in Jupyter Notebook. Notebooks can be stored as *documentation* in HTML or PDF format.

*Claims* are either inner nodes or leaves of the tree-like structure of ACs. As inner nodes, they contain a *strategy* for refinement into subclaims, as well as a reference to their *subclaims*. As leaves, claims are not further refined by subclaims and get a reference to the *evidence* supporting the claim, which are *realized blueprints*. A *conclusion* can be added as a *justification* over the evidence, i.e., if and how the evidence from one or multiple realized blueprints shows that the claim holds. A claim can contain one or multiple *contexts* or *assumptions* described in textual form. Furthermore, a claim contains a list of references to available measures. A quality **measure** together with a blueprint is intended to provide evidence for a claim. A measure can be performed in various ways. For example, a measure for detecting outliers in the dataset might either apply one selected outlier detection technique or apply multiple techniques and combine the outliers found. A **blueprint** is a concrete way to implement a measure and provides step-by-step guidance to apply it for a specific use case, which we refer to as a ***realized blueprint***. Measures contain a list of references to available blueprints that implement the measure. Blueprints contain a *justification* that they have the ability to sufficiently address the corresponding claim. Moreover, the applied blueprint keeps information on the model and data versions used, which can also be updated to create a new documentation. A *conclusion* is added to describe the contribution of the generated evidence to the corresponding claim.

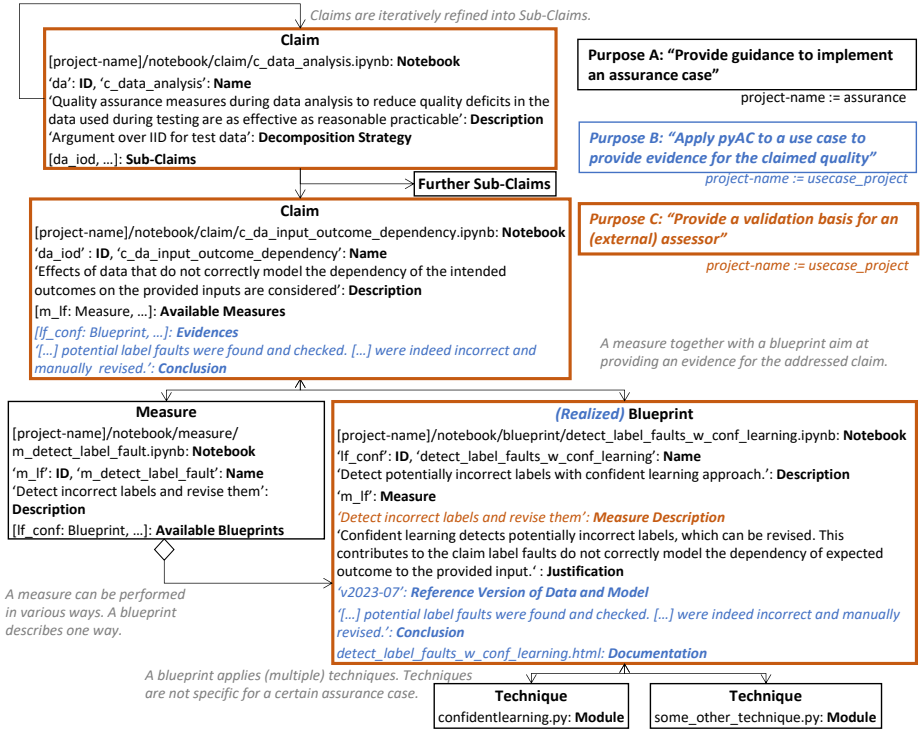

Fig. 1. Framework architecture for operationalizing ACs illustrated on the example of detecting label faults. AC elements, techniques, and their relations are depicted for the three different purposes of the framework (indicated by color).

All notebooks contain a *summary section* loading a previously stored AC element and providing summarized information on the AC element, an overview of generated documentation versions (i.e., the exported HTML/PDF versions of the notebook), and the most recently added conclusion (in the case of claims and blueprints). This section mainly targets the purpose of providing a validation basis for an assessor. The *management section* creates a new AC element, manages its versions, and stores it. Measures provide an overview of available blueprints, while claims provide an overview of available measures. Claims further reference their contributing evidence and the justifications over them. Blueprints contain an additional *blueprint section* describing the steps to generate the evidence when applying the blueprint for a specific use case.

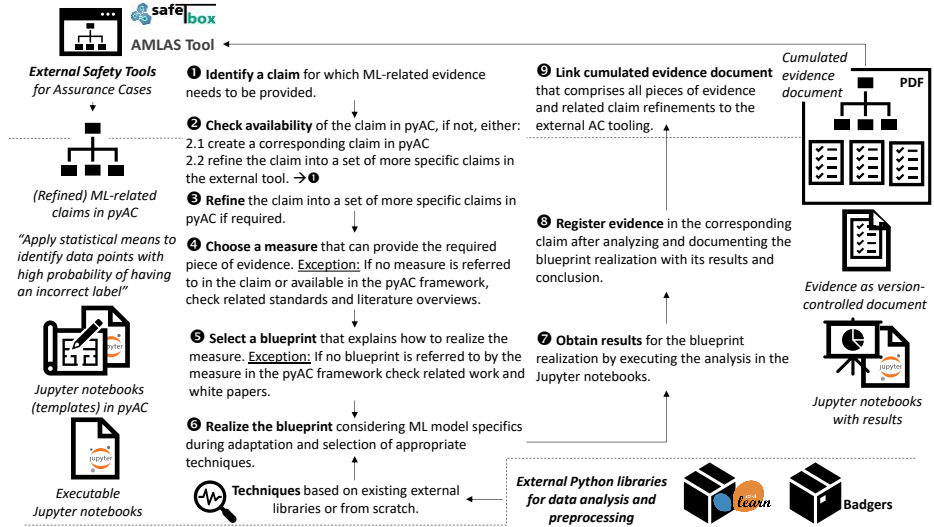

Fig. 2. Process for applying the pyAC framework for a specific use case.

Besides AC elements, the framework contains ***techniques***, which are implemented in Python modules and inherit from the class `Technique`. They are compliant with the estimator interface of the Machine Learning package scikit-learn [18]. Techniques are either based on existing packages like scikit-learn or are implemented as custom techniques. They can be used in the blueprint Jupyter notebooks; e.g., a technique using scikit-learn's isolation forest might be applied as part of a measure to detect outliers.

The framework in its bare form is intended to provide a basic collection of AC elements and techniques that are not realized and adapted for a specific use case yet. The ***process for applying the framework for a use case*** is depicted in Fig. 2. For a use case, already available AC elements can be assembled and applied. If further AC elements or techniques are needed, respective notebooks (or Python modules in the case of techniques) can be added, which will extend the framework over time. Elements and documentation are only stored when the AC elements are applied for a use case.

## 4 Quality Assurance of Test Data

In this section, we illustrate the application of the framework (using the process depicted in Fig. 2) on an excerpt of the AC that is concerned with the quality assurance of the test data. Data is used during the whole lifecycle of the ML component, e.g., for fitting the model parameters as well as for validating and testing the DDM. Hence, data quality assurance is an important part of the overall quality assurance of the ML component [3]. Compared to traditional software testing, different concepts are used for ML testing, which are mainly based on determining the performance of the ML component on a test dataset using statistical evaluation metrics [19]. Hence, deriving dependable test results strongly depends on the quality of the test data [20]. Three key quality characteristics for test data were derived from the property of random samples that they are independent and identically distributed (IID): The test data (1) was unseen during model development, (2) provides model inputs that are representative of the intended application scope of the ML component, and (3) models the relation between model inputs and intended outcomes correctly [20].

Each dataset is associated with a data lifecycle starting with the specification of requirements on the data (i.e., data specification). Data construction includes data collection and data preparation. Data analysis aims at finding weak points in the dataset in order to improve it. Data testing estimates the amount of remaining weak points to be considered in the test result of the DDM. Data operation refers to the application of the data for training, validation, or testing of the DDM. The key characteristics of test data need to be addressed by various measures during different lifecycle phases of the test data. E.g., representativity can be addressed by appropriate sampling approaches to collect data combined with approaches to enhance the data with realistically occurring quality issues (e.g., [21] [22]). A common problem regarding the correctness of the input-outcome relationship are incorrect labels, i.e., annotated ground truth information. In the following, we will illustrate the application of the presented framework on the example of detecting incorrect labels during data analysis and revising them, thereby reducing the risk arising from a test result derived from unreliable data. From

**Summary Section**
ID: lf_conf
Name: detect_label_faults_w_conf_learning
Description: Detect potentially incorrect labels with confident learning approach.
Realized measure: Detect incorrect labels and revise them
Data/model version: v2023-07
Element version: 1663591378 1690280832
Documentation version:

| Timestamp | Date and time | Data/model version | ... |
|---|---|---|---|
| 1663591392 | 2022-09-19 14:43:12 | v2022-08 | ... |
| 1690280832 | 2023-07-25 12:27:12 | v2023-07 | ... |

Conclusion: Three potential stop signs not being labeled as such were found and checked. The signs were indeed no stop signs and were not revised.

**Management Section**
Create AC element:
`ac_elem = elem_cls(**elem_param_dict)`
Set/Update version:
`ac_elem.set_element_version()`
Store AC element:
`ac_elem.save()`

Label confusion matrix (Step 3):

|  | CL-Label: 'not_stop' | CL-Label: 'stop' |
|---|---|---|
| Original label: 'not_stop' | 269518 | 3 |
| Original label: 'stop' | 11 | 6454 |

Label fault candidates (Step 4):

| Index | Original label | CL-Label |
|---|---|---|
| 326246 | No overtaking | Stop |
| 95080 | Speed limit 30 km/h | Stop |
| 216408 | No entry | Stop |

**Blueprint Section**
Step Initial: Evidence version
`ac_elem.set_documentation_timestamp() # Set evidence timestamp`
Step1: Load and prepare data
```
data=DataPreprocessor().get_dataset_df('test')  # Load test dataset
ac_elem.update_ref_version('v2023-07')  # Set data and model version
# Only differentiate between classes 'stop' and 'not_stop'
data.loc[data['signtype'] != 'stop', 'signtype'] = 'not_stop'
```
Step2: Load DDM and get prediction probabilities per class per data point
```
ddm = TrafficSignRecognitionCNN.load(path_to_ddm) # Load DDM
prop_pred = ddm.predict_proba(data) # Prediction probabilities
```
Step3: Use confident learning to compute number of label confusions per class
```
label_confusion_matrix = compute_confident_joint(data['signtype'],
    prob_pred) ➔ see label confusion matrix (left)
```
Step4: Determine data points with potentially incorrect labels, check them manually and revise
```
data_w_label_fault = data.loc[get_noise_indices(data['signtype'],
    prob_pred)] # Get indices of potentially incorrect labels
stop_sign_w_label_fault = data_w_label_fault.loc[
    data_w_label_fault['signtype'] == 'not_stop'] # Label fault
    candidates of stop signs ➔ see label fault candidates (left)
```
Step Conclusion: Describe and add conclusion, create documentation, save realized blueprint
```
ac_elem.add_conclusion('Three potential stop signs not being labeled
    as such were found and checked. The signs were indeed no stop
    signs and were not revised.') # Add conclusion
ac_elem.create_documentation() # Create HTML/PDF documentation
ac_elem.save(overwrite=True) # Save blueprint instance
```

Fig. 3. Overview of a realized blueprint notebook for dealing with incorrectly labeled data points on the example of traffic sign images annotated with their traffic sign type, illustrating the three sections and example code snippets.

a safety perspective, this contributes to the risk acceptance criterion ALARP, which states that the risk remaining after the application of the quality measures is *As Low As Reasonably Practicable* [23].

In Fig. 1, the framework architecture is illustrated on the example of detecting incorrect labels. The claim regarding quality measures applied during data analysis is based on the ALARP criterion and is divided into subclaims for each key characteristic of the test data. The quality measure for detecting and revising incorrect labels is implemented by a blueprint using confident learning [24] to identify potentially incorrect labels, which are then checked and revised if necessary. An overview of the blueprint is shown in Fig. 3, showing the summary, management, and blueprint section together with some code snippets.

## 5     Conclusion and Future Directions

We have proposed a lightweight Python-based framework named pyAC for operationalizing ACs to assure qualities of ML components, focusing on smooth integration into the daily work of data scientists. We introduced how AC claims can be refined in pyAC and how evidence supporting the subclaims is implemented by quality measures and blueprints. By applying pyAC on an example in the context of test data quality, we illustrated the process of using the framework for a use case. We further presented three main purposes of pyAC: providing templates for data scientists, instantiating them for a use case, and providing a validation basis for assessors.

We also outlined the possibilities of integrating pyAC-generated artifacts into existing AC tools that assure the software system. One possibility is for claims regarding ML components to be completely refined using the external AC tool and pyAC-generated evidence addressing the refined subclaims. Another possibility provides a higher-level claim regarding the quality of the ML component and pyAC further refines this claim in addition to generating evidence.

In future work, we plan to integrate the pyAC-generated artifacts into existing tools such as AMLAS, which already provides an argumentation structure for ML components, or safeTbox, which focuses on assuring the safety of the overall system. Integration of pyAC into external AC tools supports combining capabilities concerning data science and classical software and systems engineering.

pyAC is a step towards the assurance of AI systems, and seems promising in terms of simplifying the verification and validation of these systems by providing support and guidance for data scientists. We hope that our contribution will facilitate the development of more robust and reliable AI systems, and we look forward to further exploration and development of our framework and its potential applications.

**Acknowledgments.** Parts of this work have been funded by the German Federal Ministry of Education and Research (BMBF) in the project "DAITA", by the project "LOPAAS" as part of the internal funding program "ICON" of the Fraunhofer-Gesellschaft, by the project "AIControl" as part of the funding program "KMU akut" of the Fraunhofer-Gesellschaft, and by the German Federal Ministry for Economic Affairs and Energy in the project "SPELL".